# RECENT IMPROVEMENTS IN THE BEAM CAPTURE AT FERMILAB BOOSTER FOR HIGH INTENSITY OPERATION *

C.M. Bhat†, S. J. Chaurize, P. Derwent, M. W. Domeier, V. Grzelak, W. Pellico, J. Reid
B. A. Schupbach, C.Y. Tan, A. K. Triplett, Fermilab, Batavia IL, U.S.A.

## Abstract

The Fermilab Booster uses multi-turn beam injection with all its cavities phased such that beam sees a net zero RF voltage even when each station is at the same maximum voltage. During beam capture the RF voltage is increased slowly by using its *paraphase* system. At the end of the capture the feedback is turned on for beam acceleration. It is vital for present operations as well as during the PIP-II era that both the HLRF and LLRF systems provide the proper intended phase and RF voltage to preserve the longitudinal emittance from injection to extraction. In this paper, we describe the original architecture of the cavity phase distribution, our recent beam-based RF phase measurements, observed significant deviation in the relative phases between cavities and correction effort. Results from the improved capture for high intensity beam are also presented.

## INTRODUCTION

Fermilab has undertaken major improvements to the existing accelerators in recent years to meet the high intensity proton demands for HEP experiments both onsite as well as long baseline neutrino experiments. An important program was the "Proton Improvement Plan" (PIP) [1]. PIP had the baseline goal of extracting beam at 15 Hz from the Booster with about 4.3E12 ppp (protons per pulse). PIP completed its goals successfully in late 2016. During PIP - II [2], the plan is to increase the Booster beam intensity per cycle by >50%, and the cycle rate from 15 Hz to 20 Hz. The injection energy will be increased from 400 MeV to 800 MeV by using a newly built superconducting RF LINAC which will be completed around 2027. We have an intermediate beam intensity goal of achieving ~5E12 ppp in about two years. This will enable us to increase the beam power on the NuMI target from 700 kW to >900 kW. Concurrently, we will continue to provide 8 GeV beam to multiple low energy neutrino experiments before PIP-II comes online. The intermediate power increase helps us to better prepare Booster for the PIP-II era.

Booster is the oldest rapid cycling synchrotron (RCS) in operation in the world. Mitigation of beam losses during current operations continue to be a challenge and is of high priority. The allowed beam loss must be ≲ 475 W in the Booster ring or an average 1 W/m during its operation. Figure 1 displays a typical snapshot of the RF voltage, beam current and beam loss from its current operation.

The Booster beam cycle starts with a multi-turn H- injection scheme with no RF buckets opened. The injected beam debunches and loses its LINAC bunch structure by the end of beam injection. Next, the injected beam is captured by raising the RF voltage by changing the phase angle between two groups of RF cavities called A and B (see Fig. 3(a), 3(b). A and B have the same number of RF stations (see Fig. 4). This method of manipulating the phase is called RF *paraphase*.

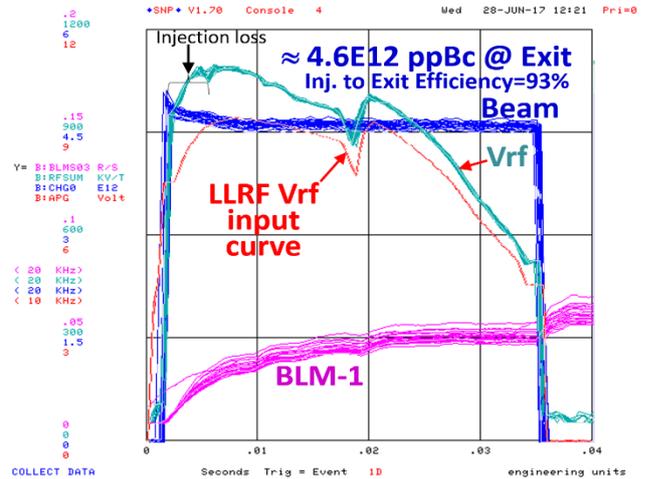

Figure 1: These are snapshots of the RF voltage, beam current, and a beam loss monitor at every 15 Hz tick.

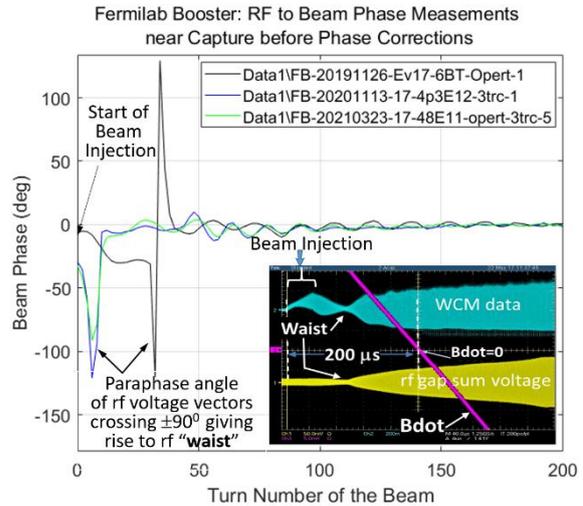

Figure 2: RF to beam phase measurements during injection and capture. The inset shows the wall current monitor (WCM) data (cyan), RF waveform (yellow), Bdot curve (magenta) during the first 400 μs of the beam cycle.

Since 2015 we have adopted an early injection scheme [3] by adding ~200 μs to the beam cycle for adiabatic beam capture. The aim is to improve the capture efficiency and reduce any beam loss arising from longitudinal beam dynamics during the first 5 ms of the beam cycle. On average, although the early injection scheme improved the longitudinal emittance by nearly 15% to the downstream machines, it did not help improve capture efficiency. Prior to our work described later in this paper, the best injection to extraction efficiency imposed a waist like structure in the RF sum voltage and in the WCM data at ~100 μs as shown in the inset of Fig. 2. The measured RF to beam phase around the waists, displayed in Fig. 2, for three different operational cases, shows sudden phase glitches in a matter of few turns. This observation contradicted results from simple longitudinal beam dynamics simulations which showed a smooth transition from injection to beam capture without any RF waist or RF phase glitch. In addition to this issue, the Booster RF voltage measured using synchrotron frequency measurements [4] was about 5% smaller than the sum of the magnitude of RF voltages from individual gap monitor outputs (Vrf shown in Fig. 1). One explanation for this reduction in the vector sum of the RF voltage is that the RF vectors from individual stations are not properly phased. These two issues led us to revisit the current configuration of the Booster HLRF and LLRF systems.

---

* Work supported by Fermi Research Alliance, LLC under Contract No. De-AC02-07CH11359 with the United States Department of Energy
† cbhat@fnal.gov.

## INJECTION SIMULATIONS

A clue for the possible causes for beam loss in the early part of the beam cycle that has an RF waist came from injection simulations described in Ref. [5]. Simulations were done using ESME [6]. The LLRF paraphase system used in the Booster during injection and beam capture is shown in Fig. 3(a) and (b) that has phase or magnitude errors. RF measurements during injection indicated that the observed waist in the WCM data might arise from the $V_a$ and $V_b$ vectors (dashed lines in Fig. 3(a)) not being anti-parallel. $V_a$ and $V_b$ are the vector sums of the RF voltages of all the cavities in groups A and B respectively. In the current operations, we have ten RF cavities in each group. Figure 3(b) shows an example of a phase error between cavity voltage vectors. These errors give rise to emittance growths and possible beam particle losses. The simulation results for an ideal case with $|\theta_a|$ = $|\theta_b|$ = 90 deg during injection and $|\theta_a|$ = $|\theta_b|$ at the end of beam capture with $|V_a|$ = $|V_b|$ are shown in Fig. 3(c) and 3(d). This illustrates a case with no emittance growth. If $|V_a/V_b| \neq 1$ and or, one of the paraphase angle >90 deg at injection, then as the paraphase angle starts changing during the beam capture process, the magnitude of the RF voltage sum vector reaches a minimum value resulting in beam acceleration or deceleration along with debunching. In the illustration shown here, the debunching took place in ~20 μs which is much faster than the synchrotron period at that time. This led to >30% emittance dilution. Figures. 3(e) and 3(f) show the phase space distributions from the end of injection to the end of beam capture, respectively. Further simulations showed that if cavities in groups A and/or B have phase errors or magnitude differences (as shown in Fig. 3(b)) then there is emittance growth even without any RF waist. Therefore, it is quite important to ensure that the cavities have the proper phase and have comparable gap voltages that are within a few percent.

## RF PHASE MEASUREMENTS

The Booster uses 20 RF cavities in operation (with two additional cavities as backup), each with a gap voltage of 50 kV. During beam acceleration, the RF frequency ramps from 37.92 MHz to 52.81 MHz. Figure 4 shows a schematic of the RF cavity phasing diagram and arrangement of group A and B cavities in the Booster. Theoretically, the closest pair of A and B cavities must be $\beta\lambda$ apart for acceleration. But due to space constraints, they are closer to $\beta\lambda/2$. The solution to this problem is for the LLRF to send RF to groups A and B that has a phase difference of 180 deg plus phase compensation that ensures that the A and B cavities are exactly $\beta\lambda/2$ (+ $n\beta\lambda$) apart. For example, the arrangement and spacing of four consecutive cavities are shown in Fig. 5. The center-to-center distance between two adjacent cavities in a sector (sectors 14 and 15 as shown in Fig. 4) is 2.385 m that is $<\beta\lambda/2 = 2.816$ m. Between any two sectors the spacing is $n\beta\lambda + \Delta\theta\beta\lambda/2\pi$. The quantity $\Delta\theta$ =28 deg, is the required phase compensation because the cavity separation is not exactly an integer multiple of $\beta\lambda/2$. The fanout phase shifter (also called a *brass box*) shown in Fig. 5 adds the required phase compensation provided the connecting cables from each cavity to the LLRF box are identical.

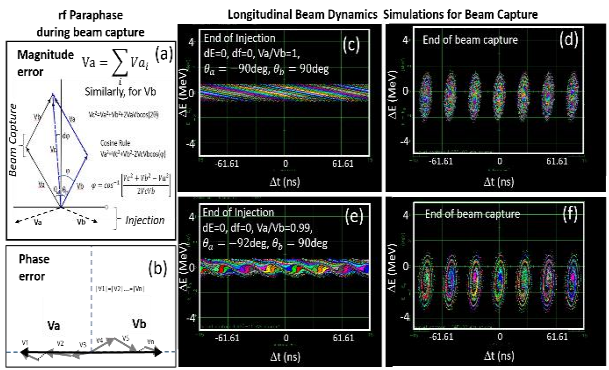

Figure 3: (a) and (b) are schematics of two scenarios of paraphase RF voltage vectors during beam injection into Booster. (c) - (f) are simulated ($\Delta E, \Delta t$)-phase space distributions for these two scenarios of paraphase.

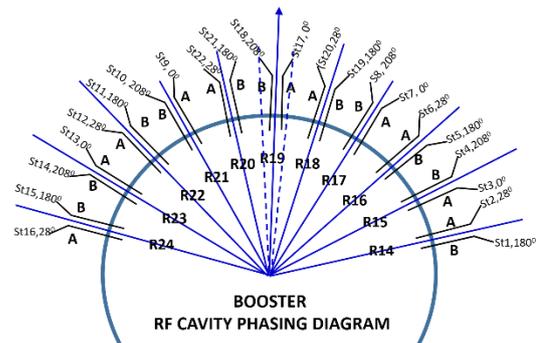

Figure 4: Schematic of the Booster RF cavity configuration for 22 RF cavities with added phases between cavities.

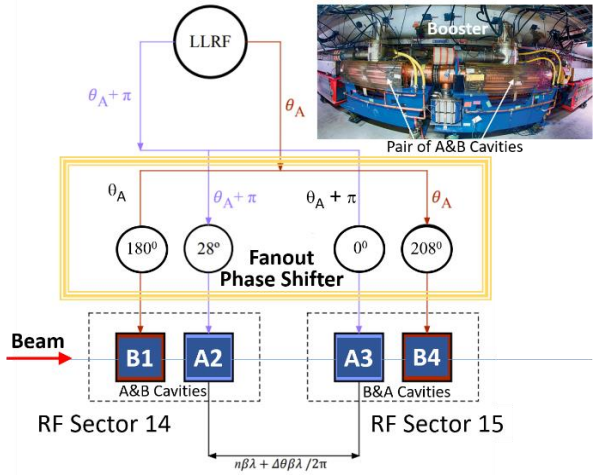

Figure 5: Schematic of Booster RF cavity configuration with fanout phase shifter for a set of four cavities.

To measure the relative phase between any pair of RF cavities, we performed the following: i) Booster was put into a non-ramping state at the injection energy of 400 MeV with the RF frequency held constant, ii) in the paraphase module, the paraphase angle between A and B RF stations were set to 180 deg. In principle, if the system is perfect, A and B RF voltage vectors should be anti-parallel, iii) the magnitude of the RF voltage vector for each station is set to ~30 kV (within about 2%). We chose one cavity in group A to be the master station and then used beam to measure the phase of each cavity in group B. We expected that if A and B was perfectly anti-parallel, then there is no bunching, i.e., there is no RF component (37.92 MHz) carried by the beam. In practice, this was not the case because we see this component. We could measure the phase error between these two cavities by simply knobbing the paraphase angle until the RF component carried by the beam vanishes.

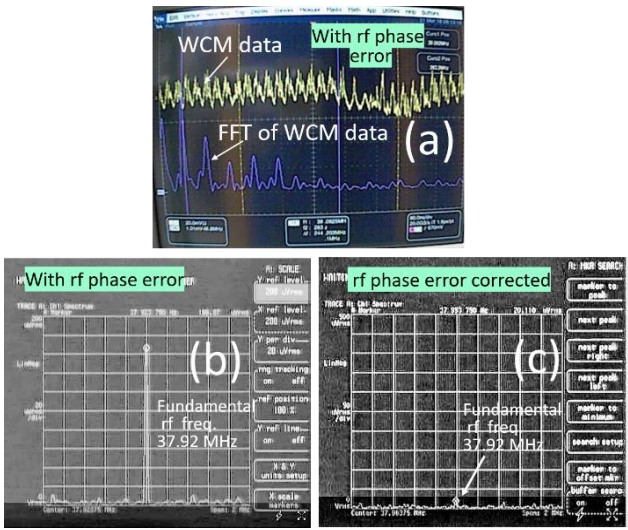

Figure 6: (a) A typical time domain and (b), (c) VSA data of the beam before and after phase correction.

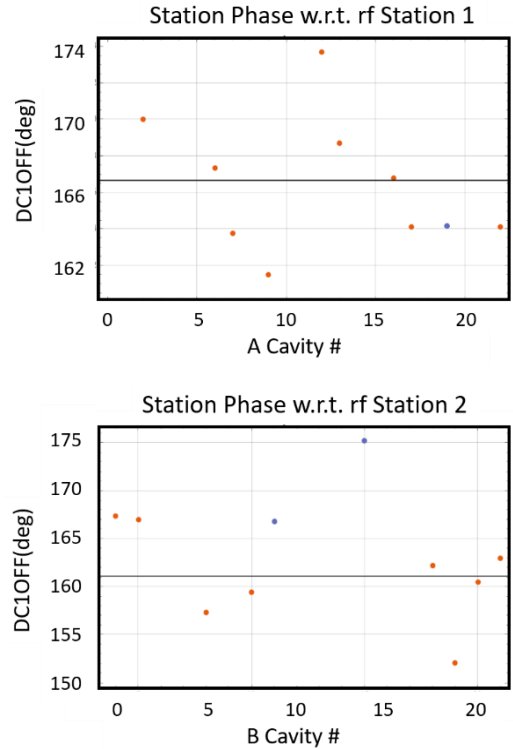

Figure 7: (top)The measured phase differential offset of all group A cavities relative to station-1 (group B cavity) and (bottom) similar measurement of all group B cavities relative to station 2 (group A cavity) before phase correction.

A typical WCM data and its FFT measured with a Tektronix, TDS7154B Digital Phosphor about 1 ms after injection that shows bunching is shown in Figure 6(a). The WCM data measured with a vector signal analyzer (VSA) is shown in Fig. 6(b).

Since we can see the 37.92 MHz component prominently in Fig. 6(b), this indicates that these two RF cavities are not anti-parallel. To make them anti-parallel, we adjusted a phase differential offset device DC1OFF in the LLRF system until the 37.92 MHz component disappears. Figure 6 (c) shows the VSA data after adjustment. The value (DC1OFF -180) deg when this component disappears is the phase error between these pair of cavities. After working on this pair, the same method was used to measure DC1OFF for the remaining pairs. The as found measured DC1OFF are shown in Fig. 7. The average DC1OFF before any correction was 166.7 deg between group A stations to a reference group B station (station 1). And 161.1 deg between group B stations to a reference group A station (station 2). If there was no phase error, we expected DC1OFF to be 180 deg.

The observed systematic phase error can come from a) the LLRF system, b) from the mismatch of cable lengths from the LLRF system to fanout phase shifter box, fanout phase shifter box to the cavities or from cavities to the fanback sum box (the last one will not affect the observed phase error, but has an effect on the measured Vrf), c) the circuits inside the fanout phase shifter box or d) from the improper cavity spacing in the Booster ring. Or from all the above. We looked at each of these items as a possible source of the error but ruled them all out except for the Booster LLRF. We found that the paraphase module in the LLRF was the major source of the error because of an incorrect cable length. We fixed the problem and remeasured DC1OFF after correction.

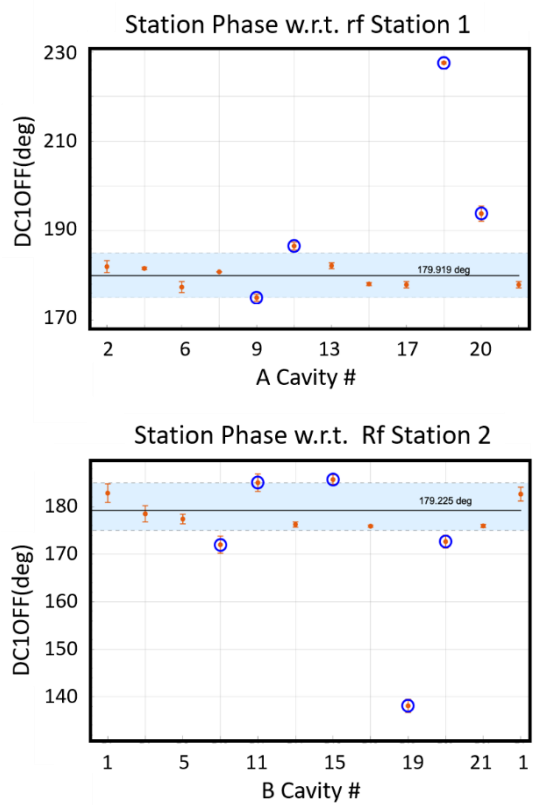

Figure 8: Phase measurement after correction. Figure description is same as that for Fig. 7.

Figure 8 displays measurement results for all the cavities after the phase error was corrected. We found that the average DC1OFF is 179.9 deg and 179.2 deg for group A stations relative to the reference group B station (station-1) and group B stations relative to the reference group A station (station-2), respectively. This met our specification of < 5 deg from 180 deg. Stations 19 and 20 were outliers during these measurements, and they were corrected later. These two stations will need further beam-based phase measurements to verify their corrections.

## OBSERVED IMPROVEMENTS

After correction, the measured RF to beam phase during injection and beam capture is shown in Fig. 9 for an extraction intensity of 4.5E12 ppp with about 94% injection to extraction efficiency. This is the normal intensity for 15 Hz operation. The measurement results look like our simulation results for a properly corrected RF system. There was also a noticeable improvement in the capture process: under low beam repetition conditions, we had up to about 5.7E12 ppp at extraction with a comparable injection to extraction efficiencies. Currently we are optimizing the Booster for higher intensity as per users' demand.

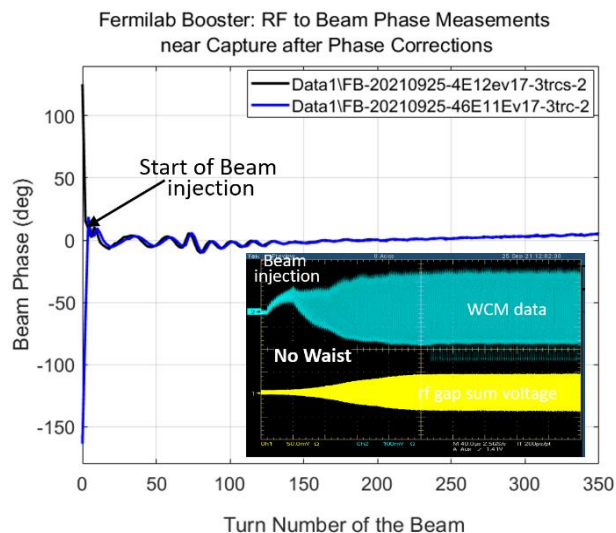

Figure 9: The RF to beam phase measurements during injection and capture after all RF cavities have been phased correctly. The shown data is for 4.5E12 ppp in the Fermilab Booster.

## ACKNOWLEDGEMENTS

Special thanks due to RF personnel and MCR crew for their help during these measurements.